\documentclass[conference]{IEEEtran}

\usepackage[T1]{fontenc}

\usepackage{amsmath, amsthm, amsfonts, amssymb, amsbsy}
\usepackage{setspace}
\usepackage{graphicx}
\usepackage{pstricks,arydshln}
\usepackage{algpseudocode}
\usepackage{algorithm}
\usepackage{multirow}
\usepackage{subfigure}
\usepackage{verbatim}
\usepackage{epstopdf}

\newtheorem{theorem}{Theorem}
\newtheorem{lemma}{Lemma}

\newcommand{\Lref}[1]{Lem\-ma\,\ref{#1}}
\newcommand{\Cref}[1]{Co\-ro\-lla\-ry\,\ref{#1}}

\newcommand\GF{{\mbox{GF}}}

\newcommand{\cC}{{\cal C}}
\newcommand{\cP}{{\cal P}}

\begin{document}

%
\title{\huge HARQ Rate-Compatible Polar Codes for Wireless Channels}

\author{\IEEEauthorblockN{Mostafa El-Khamy$^\diamond$, Hsien-Ping Lin$^\circ$, Jungwon Lee$^\diamond$, Hessam Mahdavifar$^\diamond$, Inyup Kang$^\diamond$ }
\IEEEauthorblockA{
$^\diamond$Modem Systems R\&D, Samsung Electronics, San Diego, CA 92121, USA\\
$^\circ$Department of Electrical and Computer Engineering, University of California, Davis,
CA 95616, USA
\\{\small Email:  $^\diamond$\{mostafa.e,  jungwon2.lee, h.mahdavifar, inyup.kang\}@samsung.com}, $^\circ$hsplin@ucdavis.edu
 }}

\renewcommand\figurename{Fig.}
\maketitle
\begin{abstract}

A design of rate-compatible polar codes suitable for HARQ communications is proposed in this paper. An important feature of the proposed design is that the puncturing order is chosen with low complexity on a base code of short length, which is then further polarized to the desired length. A practical rate-matching system that has the flexibility to choose any desired rate through puncturing or repetition while preserving the polarization is suggested. The proposed rate-matching system is combined with channel interleaving and a bit-mapping procedure that preserves the polarization of the rate-compatible polar code family over bit-interleaved coded modulation systems. Simulation  results on AWGN and fast fading channels with different modulation orders show the robustness of the proposed rate-compatible polar code in both Chase combining and incremental redundancy HARQ communications.

\end{abstract}

\section{Introduction}

Polar codes, introduced by Ar{\i}kan, are the first
class of error-correcting codes that provably achieve the symmetric capacity of memoryless channels with low-complexity encoders and decoders \cite{Arikan_09}.
Ar{\i}kan applied a linear transformation over $\GF(2)$ on an input vector of length $N=2^n$ bits based on the $n$-th Kronecker power $F^{\otimes n}$ of the $2\times 2$ matrix
\begin{equation} \label{Eq_G}
F=\left[\begin{array}{cc}
1 & 0 \\
1 & 1 \end{array}\right],
\end{equation}
which is referred to as a \emph{kernel matrix}.
As the number of polarization levels $n$ increases, the input channels are transformed into bit-channels that tend to become either noiseless or completely noisy channels under \emph{successive cancellation (SC) decoding}. The fraction of the noiseless bit-channels approaches the symmetric capacity of the transmission channel
\cite{Arikan_09}.
The encoding and SC decoding complexities of Ar{\i}kan's polar codes are relatively low at $N \log_2 N$.

To design hybrid automatic repeat request (HARQ) rate-compatible (RC) codes for wireless systems, it is required that the information block length is fixed across all codes with different rates.  It is also required that the transmitted bits can be flexibly chosen according to the desired code rate, and HARQ scheme, Chase combining (CC) or incremental redundancy (IR). Although there have been attempts to construct codes with different rates or lengths, they do not satisfy the flexible HARQ design requirements. The block lengths of Ar{\i}kan's polar codes were restricted to a power of two.
Polar codes with arbitrary length can be constructed by replacing the kernel matrix $F$ in (\ref{Eq_G}) with a non-singular binary matrix of size $\ell\times \ell$  \cite{Korada_Sasoglu_Urbanke_10}.
However, adjusting the polar code block lengths through changing the size of the kernel matrix implies different pairs of encoders and decoders for each block length, as well as increased decoder complexity at larger $\ell$. Puncturing polar codes to obtain length-compatible codes has been viewed as a process of reduction of the size of the matrix $F^{\otimes n}$, where the reduced matrix with the largest polarizing exponent is found by exhaustive search for each desired rate \cite{matrix_reduction_polar}.
Changing the code rate through information shortening and random  puncturing of the output bits were investigated \cite{first_com_polar}. Another scheme changes the transmission rate through quasi-uniform puncturing as well transmissions of the input information bits \cite{harq_polar}.
Several criteria based on minimizing the error probability have been investigated to chose a puncturing pattern
\cite{punc_pattern_polar}, where the search criterion is tested for each possible puncturing pattern, at each desired code length and desired code rate.

One disadvantage of the previous algorithms is that they do not necessarily result in a HARQ RC family of codes. Moreover, the puncturing pattern has to be exhaustively found at each code length and code rate.
In this paper, we propose an efficient algorithm that finds the best order for puncturing the output bits on a base polar code, while guaranteeing that the codes with different rates are nested.
Our simulation results show that this scheme achieves a decoding performance similar to that when the optimal puncturing pattern is selected for a given rate and block length.
We show how a regular puncturing pattern of a longer polar code can be derived from the puncturing pattern of the base polar code, while preserving the code polarization, based on the compound polar code construction \cite{compound_polar}. This reduces the complexity of the code design, where we do not need to exhaustively search for the puncturing at each desired code rate and code length. Moreover, the time and space complexities of operation at both the transmitter and the receiver are reduced.  We propose a simple rate-matching scheme which implements the compound punctured polar codes and allows for flexible puncturing or repetition of code bits. Our scheme supports both HARQ Chase combining, when the bits of the retransmissions are the same as those of the first transmission, and HARQ incremental redundancy, when the retransmissions constitute some coded (redundancy) bits that have not been transmitted before and may also constitute some of the previously transmitted bits. We show how the proposed rate-matching structure also provides channel interleaving and bit-mapping for bit-interleaved coded modulation (BICM) while maximizing code polarization.

The rest of the paper is organized as follows. Preliminaries of polar codes and estimation of their bit-channel error probability are reviewed in Section \ref{Pre}.
In Section \ref{Sec_rc_polar}, we describe the different aspects for the design of the proposed rate-compatible polar codes.
Analysis of the achievable rates and the decoding performances of the proposed rate-compatible polar codes with different HARQ schemes are presented in Section \ref{Sec_simulation}.
Conclusions are made in Section \ref{conclude}.

\section{Preliminaries} \label{Pre}
The polar code construction of length $N=2^n$ is based on the $n$-th Kronecker power $F^{\otimes n}$ of the $2\times2$ matrix in (\ref{Eq_G}).
Assume the information sequence is denoted as $u_1^N=(u_1,u_2,\ldots,u_N)$ and the corresponding codeword is denoted as $x_1^N=(x_1,x_2,\ldots,x_N)$, where $x_i$, $u_i\in\{0,1\}$ for $1\leq i\leq N$.
The encoding of a polar code can be described by $x_1^N=u_1^NB_NF_2^{\otimes n},$ where $B_N$ is a bit-reversal permutation matrix \cite{Arikan_09}.
$x_1^N$ is transmitted through $N$ independent copies of a binary input discrete memoryless channel. The received sequence is denoted as $y_1^N=(y_1,y_2,\ldots,y_N)$.
The combined channel, $W_N$, between $u_1^N$ and $y_1^N$ is described through the transitional probabilities
\begin{equation}
W_N(y_1^N|u_1^N) \triangleq P(y_1^N|x_1^N)=\prod_{i=1}^N W(y_i|x_i).
\end{equation}

Through  recursive \emph{channel splitting} and \emph{channel combining} operations, a bit-channel for each input information bit $u_i$ is defined as \cite{Arikan_09}
\begin{equation}
\begin{split}
W_N^{(i)}(y_1^N, u_1^{i-1}|u_i)
\triangleq P(y_1^N, u_1^{i-1}|u_i)\\
=\hspace{-5mm}\sum_{u_{i+1}^N\in{\{0,1\}}^{N-i}}\frac{1}{2^{N-1}}W_N(y_1^N|u_1^N).
\end{split}
\end{equation}
Based on these bit-channel models, Ar{\i}kan proposed a recursive SC decoding algorithm \cite{Arikan_09}.

\begin{figure}[]
\begin{center}
\includegraphics[width=0.2 \textwidth ]{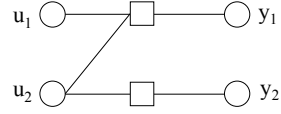}
\end{center}
\caption{Basic encoding and decoding structure of Polar codes
 \label{basic}}
\vspace{-0.2cm}
\end{figure}

For uniform channel input, the  error probability (EP) of the polarized bit-channels under genie-aided SC decoding on erasure channels can be recursively calculated through their Bhattacharrya parameters \cite{Arikan_09}. For other channels, the bit-channel error probabilities can be estimated through Monte-Carlo simulations of the genie-aided SC decoder. In case of SC decoding on AWGN channels, the bit-channel EPs  can be approximated with reasonable accuracy by density evolution and approximating the outputs at each SC decoding step with Gaussian random variables \cite{GA_polar}.
Assume the all-zero codeword is transmitted and the variance of the AWGN channel is $\sigma^2$.
We denote the logarithmic likelihood ratio of the received $y_i$ as $L(y_i)$, which is assumed to have probability distribution ${\cal N}(\frac{2}{\sigma^2},\frac{4}{\sigma^2})$.
From the basic encoding and decoding structure of polar codes in Fig.\ref{basic}, we can
find $\textbf{E}(L(u_1))$ and $\textbf{E}(L(u_2))$ from $L(y_1)$ and $L(y_2)$, where \textbf{E} denotes the expectation. The error probability of bit-channel $u_i$ is estimated by \cite{GA_polar}
\begin{equation}P(\mathcal{E}_i)=Q(\sqrt{\textbf{E}[L(u_i)/2]} )). \end{equation}
 With different $L(y_1)$ and $L(y_2)$, the Gaussian approximation (GA) can be written as
\begin{equation}
\begin{split}
\textbf{E}[L(u_1)]
=\phi^{-1}(1-(1-\phi(\textbf{E}(L(y_1))]))(1-\phi(\textbf{E}(L(y_2))])))\label{ga_1} \\
\end{split}
\end{equation}
\begin{equation}
\textbf{E}[L(u_2)]
=\textbf{E}[L(y_1)]+\textbf{E}[L(y_2)],
\label{ga_2}
\end{equation}
where
\[ \phi(x)= \left\{
 \begin{array}{ll}
 1-\frac{1}{\sqrt{4\pi x}} \int_{-\infty}^\infty \text{tanh}(\frac{u}{2})e^{-\frac{(u-x)^2}{4x}}du, & \mbox{if} \ x >0,\\
 1, & \mbox{if} \ x=0.
 \end{array}\right. \]

Through recursively applying (\ref{ga_1}) and (\ref{ga_2}) on the basic  encoding and decoding structure of a polar code,  $\textbf{E}(L(u_i))$ of each information bit $u_i$ can be calculated from the received $L(y_i)$ for $1\leq i\leq N$. In case an output channel is punctured, then its variance is set to infinity.

\section{Proposed rate-compatible polar codes}\label{Sec_rc_polar}

In this section, we describe the different aspects of the proposed HARQ rate-compatible polar code design.

Assuming an $(N,k)$ mother polar code of rate $k/N$, where $k$ is the cardinality of the information set $\mathcal{I}$.
 $\mathcal{I}$ can be chosen to constitute of the $k$ bit-channels with the lowest estimated bit-channel error probabilities
to minimize the union bound on the decoder block error rate (BLER)  \cite{Arikan_09},
\begin{equation}
P(\mathcal{E}) \leq \sum_{i \in \mathcal{I}} P(\mathcal{E}_i). \label{eq:PE}
\end{equation}

Our goal is to design a family of nested RC codes, with rates $k/N$ or higher as in $\{ k/N, k/(N-1), \ldots, k/(N-m)\}$ by puncturing coded bits, as well as below $k/N$ by repeating some of the coded bits at the same transmission.  Also, the transmission rate can be allowed to go above $1$ by transmitting less than $k$ bits if $N-m < k$. In this design, the polar encoding is not constrained to be systematic.

\subsection{Progressive puncturing algorithm}
Let the desired punctured code rate be $k/(N-m)$. The exhaustive search approach \cite{punc_pattern_polar} estimates the information set EP (\ref{eq:PE}) after puncturing $m$ bits, for all $N \choose m$ possible puncturing patterns, and chooses the one that minimizes (\ref{eq:PE}). Alternatively, for different applications, the design criterion (\ref{eq:PE}) can be changed. Whereas exhaustive search will find the optimal puncturing pattern for that rate, it does not guarantee that the puncturing patterns of different rates are nested.

To address this, we propose an efficient greedy algorithm to find the best puncturing patterns for RC polar codes.
The \emph{Progressive Puncturing Algorithm} (PPA) finds the puncturing pattern that punctures $m+1$ bits such that this puncturing pattern minimizes the design criteria (\ref{eq:PE}) and is constrained to contain the  $m$ previously punctured bits. Hence, it results in a nested family of codes, where the code of rate $k/(N-m+1)$ is obtained from the code of rate $k/(N-m)$ by transmitting the $m$th punctured bit.  The set of indices of the punctured bits is denoted by $\cP$ which is also referred to as the puncturing pattern. Given a certain puncturing pattern $\cP$ of size $m$, the $(m+1)$th punctured bit is selected from the remaining $N-m$ non-punctured coded bits by testing the design criterion $N-m$ times. Hence, to find the puncturing order for $1 \leq m \leq N$, the design criterion needs to be tested $N(N+1)/2$ for the PPA, which is significantly less than the $2^N$ searches  required with exhaustive search.

\begin{figure}[]
\begin{center}
\includegraphics[height=2.4 in, width=0.45 \textwidth ]{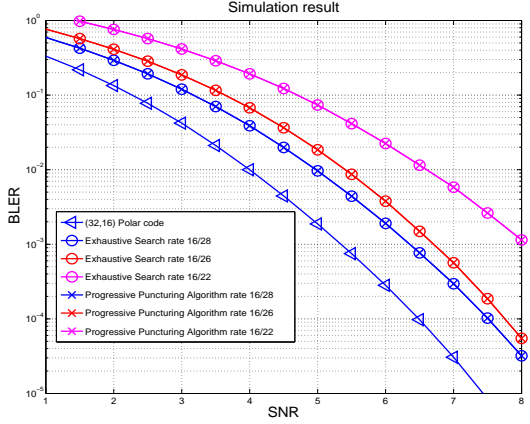}
\end{center}
\caption{Progressive versus exhaustive search puncturing of polar codes   \label{fig:PPA}}
\vspace{-0.2cm}
\end{figure}

In Fig. \ref{fig:PPA}, the decoder BLER is estimated using  (\ref{eq:PE}), where the bit-channel EP is found by the GA on an AWGN channel after code puncturing. Assuming an $(32,16)$ polar code designed at $\mbox{SNR}=3$ dB, the performance of the punctured polar code by PPA is shown to overlap with that of the punctured polar code found by exhaustive search, for $m\in \{0,4,6,10\}$.

\subsection{Two-step polarization and regular puncturing \label{Sec_regular}}

Although, the PPA has significantly smaller complexity than exhaustive search, it is not practical to repeat it for every code length and rate. Moreover, even the PPA algorithm becomes infeasible at very large block lengths as it requires estimating the bit-channel error probabilities.
In our proposed scheme, the PPA is only run on a base polar code of short length, e.g. $N'=32$.
The base polar code is then further polarized into the polar code with the desired length $N$. This two-step polarization approach results in a regular puncturing pattern on the longer polar code of length $N$, which is derived from the puncturing sequence of the base code of length $N'$. Another advantage of the two-step polarization approach is that it avoids storing puncturing patterns for each code length and code rate.

\begin{figure}
\begin{center}
\includegraphics[width=0.4 \textwidth ]{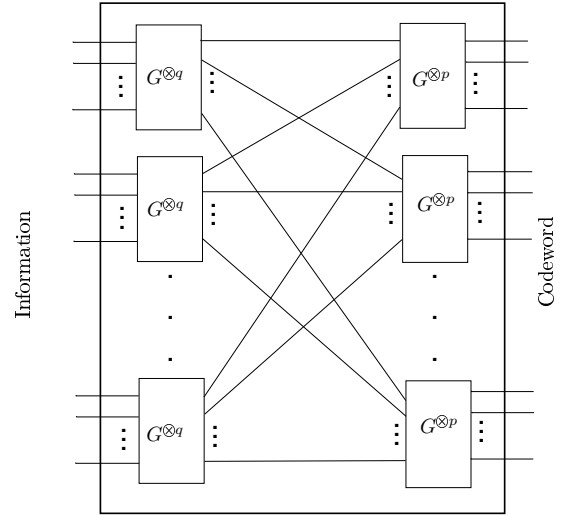}%
\end{center}
\caption{Encoding scheme based on $B_{2^{n}}(F_2^{\otimes p}\otimes F_2^{\otimes q})$\label{5kron7_g}}
\vspace{-0.2cm}
\end{figure}

Consider an $(N,k)$ mother polar code of length $N=2^n$ and  a base polar code of length $N'=2^p$.
The encoding structure with generator matrix $B_{2^{n}}(F_2^{\otimes n})$ can be decomposed into two stages as
$B_{2^{n}}(F_2^{\otimes p}\otimes F_2^{\otimes q})$.
The first encoding stage consists of $2^p$ polar code encoders of length $2^q$, and the second encoding stage
consists of $2^q$ polar code encoders of length $2^p$. This is shown in Fig. \ref{5kron7_g},
where we denote the polar encoding structures $B_{2^{p}}(F_2^{\otimes p})$ and  $B_{2^q}(F_2^{\otimes q})$ by $G^{\otimes p}$ and $G^{\otimes q}$, respectively.
The PPA is run on the base code of length $2^p$ to find the puncturing sequence. To obtain a code rate of $k/(N-m2^q)$, the first $m$ bits in the progressive puncturing sequence are punctured at the output of each $G^{\otimes p}$ code at the second encoding stage.
This results in a regular puncturing pattern on the long code, derived from base code's progressive puncturing pattern.
This regular puncturing pattern results in having the output channels of each $G^{\otimes q}$ code be transmitted on multiplicities of the same channel type, either punctured channels with zero capacity or the physical transmission channels. Hence, this construction is a generalization of the compound polar code construction  \cite{compound_polar}, and corresponds to
an  $2^p$ multi-channel compound polar code, which we will show is beneficial when mapping the bits to   symbols of higher order modulations.
To satisfy HARQ requirements, the information set is fixed across all codes with different rates in the  same family with mother code length $N$, and is chosen to minimize the BLER of a high rate code in the RC family. Besides having the best performance at high code rates, selecting the information set at the high-rate code guarantees that no zero-capacity channels exist in the information set of the lower-rate codes.

\subsection{Rate-matching, channel interleaving, and bit-mapping}\label{Sec_interleaver}

In Section \ref{Sec_regular}, we proposed an approach to obtain a regular puncturing pattern of a large block length based on the puncturing pattern of smaller length selected by the progressive puncturing algorithm.
However, with this regular puncturing structure, the punctured codeword length is restricted to $\ell 2^q$  and the transmission code rate is also restricted to $k/(\ell 2^q)$, for $\ell \in \{0,1,\dots, 2^p\}$.
To achieve a finer resolution of code rates, we propose a simple rate-matching scheme that preserves the regular puncturing obtained by the two step polarization and the puncturing order on the base code found by the PPA.

For a polar code of block length $N=2^n$, we follow the approach in Section \ref{Sec_regular} and consider the encoding structure with two stages based on $B_{2^{n}}(F_2^{\otimes p}\otimes F_2^{\otimes q})$, where $p$ and $q$ are positive integers and $p+q=n$.
First, the $2^n$ coded bits, $(x_1,x_2,\ldots,x_N)$, are arranged into a $2^{q}\times2^p$ matrix such that the $(i,j)$th element of this matrix is $x_{(i-1)2^p+j}$, for $1\leqslant i \leqslant 2^{q}$ and $1\leqslant j \leqslant 2^{p}$.
The first row of this matrix is $(x_1,x_2,\ldots,x_{2^p})$.
In this structure, each row corresponds to the component $G^{\otimes p}$ code, and each column corresponds to the component
$G^{\otimes q}$ code.
Next, the $2^p$ columns are permuted according to the reverse of the puncturing order found by the PPA on the base code of length $2^p$. To be more specific, we rearrange the columns such that the codeword bits located in the last $m$ columns of the permuted matrix are the punctured bits of the regular puncturing pattern which punctures $m$ bits in every $2^p$ bits, for $m\in\{1,2,\ldots,2^p\}$.
This rate-matching structure inherently does row-column channel interleaving to the polar codeword, where the transmitted bits are written row-wise into the matrix and read column-wise after the described column permutation.
Let the desired transmission rate be $k/L$. The transmitted codeword bits, denoted as $(\hat{x}_1,\hat{x}_2,\ldots,\hat{x}_L)$, are read from the permuted matrix such that $\hat{x}_{(j-1)2^q+i}$ is the $(i,j)$th element of the permuted matrix. 
 If $L=N-m2^q$, then the last $m$ columns will not be transmitted, and the  two-step compound polarization with regular puncturing is preserved. Moreover, this rate-matching structure does not constrain $L$  to be a multiple of $2^q$, where the first $L$ bits are always read column-wise, after column interleaving. From the PPA, the column interleaving guarantees that the non-transmitted (punctured) bits will have the least impact on the BLER given by (\ref{eq:PE}). Moreover, if $L > N$, the matrix is treated as a circular array, where the $(N+1)$th transmitted bit is the $(1,1)$th bit of the array, while reading column-wise. According to the application, further polynomial interleaving can be applied on the output interleaved sequence.

\begin{figure}
\begin{center}
\includegraphics[width=0.45 \textwidth ]{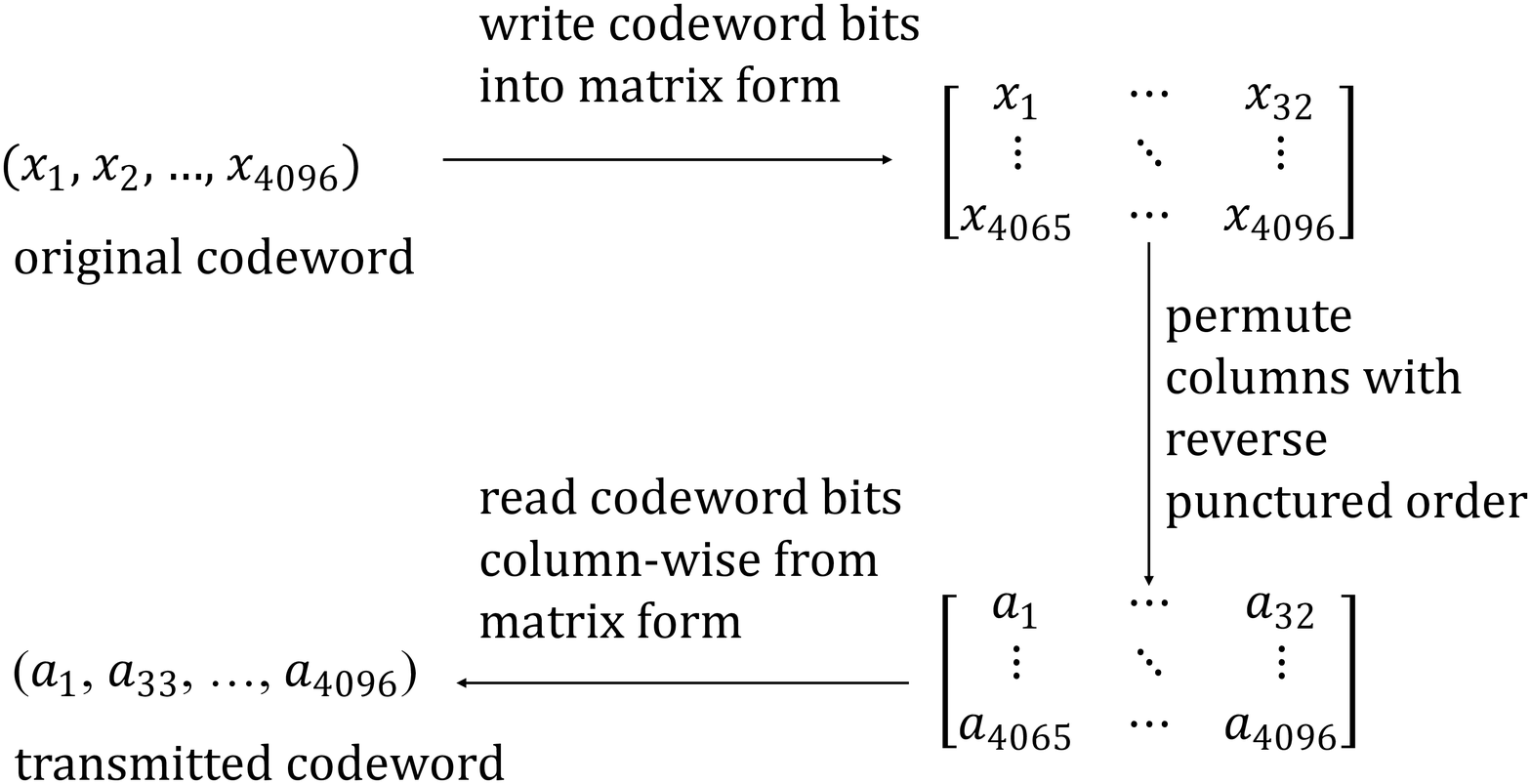}
\end{center}
\caption{Example of rearrangement of codeword of length $2^{12}$ with the rate-matching structure \label{interleaver}}
\vspace{-0.2cm}
\end{figure}

For a codeword of length $2^{12}$,  the rate-matching structure is depicted in Fig.\ref{interleaver}.
In this example, we consider the encoding structure based on $B_{2^{12}}(F_2^{\otimes 5}\otimes F_2^{\otimes 7})$.
The codeword of length $2^{12}$, $(x_1,x_2,\ldots,x_{4096})$, is arranged into an $2^{7}\times2^5$ array. Then, the columns of the array are permuted according to the reverse of the puncturing order found by the progressive puncturing algorithm.
The transmitted codeword is read from this permuted array column-wise.

\begin{figure}[t]
\begin{center}
\includegraphics[width=0.48 \textwidth ]{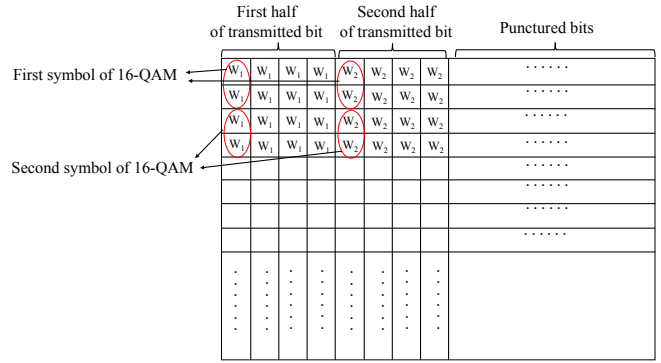}
\end{center}
\caption{Bit-mapping from codeword bits to 16-QAM symbols in matrix form \label{bit_mapping}}
\vspace{-0.2cm}
\end{figure}

In case bit-interleaved coded modulation (BICM), this structure can be combined with  bit-mapping to higher-order modulated symbols, such as 16-QAM or 64-QAM. We apply a particular bit-mapping scheme under this rate-matching structure to satisfy the compound polarization structure.  The transmission channel with $M$-QAM modulation  and gray mapping can be modeled as a multi-channel with $\log_2(M)/2$ different channel types \cite{compound_polar}, where BICM detection results in $\log_2(M)/2$ BICM channels of different reliability.
The proposed bit-mapping is based on our generalized compound polar code structure, where the $2^p$ multi-channels of the base polar code are mapped to the  $\log_2(M)/2$ BICM channels and the zero-capacity channels resulting from puncturing.

An example for the bit-mapping from the binary coded bits to the $16$-QAM symbols is shown  in Fig. \ref{bit_mapping}, where the two constituent BICM channels from the $16$-QAM modulation are denoted as $W_1$ and $W_2$.
This is combined with rate-matching, where the the first half of transmitted bits are transmitted on the first constituent channel and the second half are transmitted on the other constituent channel.

Punctured polar coding with $64$-QAM modulation and gray mapping can be also modeled as polar coding over a multi-channel with four different channel types, three channels with different reliabilities are due to $64$-QAM BICM with gray-mapping and the fourth channel is  the zero-capacity channel due to puncturing.
After rate-matching, the transmitted columns are grouped into three groups of the same size and each mapped to a different BICM channel type. The idea in this bit-to-symbol mapping is to guarantee, whenever possible, that each component $G^{\otimes q}$ polarizes over the same channel type. Hence, different assignments of the columns to the BICM bit-channels are possible.

\subsection{HARQ Chase combining and incremental redundancy}\label{Sec_HARQ}

The proposed rate-matching scheme allows transmission of rate-compatible polar codes with both Chase combing (CC) and incremental redundancy (IR).
If the decoding result of the previous transmission failed, then  the same modulated symbols are re-transmitted with CC, but different symbols can be transmitted with IR.
Assume the transmission rate of $k/L$ is fixed across all HARQ transmissions and a maximum of $t$ transmissions.
The proposed structure allows seamless IR or Chase transmissions at any rate. With HARQ CC, all transmissions will send the $L$ bits column-wise from the permuted matrix, starting from the first column. With IR, the $r$th transmission may start sending the $L$ bits column-wise starting from $((r-1)2^p/t + 1)$th column. The BICM bit-mapping is then applied by circularly shifting the BICM column assignments  $(r-1)2^p/t$ columns.
Other combinations of the IR column assignments of each transmission with the BICM bit-channel assignments that preserve the two-step polarization structure are also possible.

\section{Analysis of proposed RC polar codes}\label{Sec_simulation}
First, we provide a theoretical analysis on the achievable rates, and then we provide numerical simulation results.

Consider the modified polar transformation of Fig. \ref{5kron7_g} with base polarization block $G^{\otimes p}$, $N' = 2^p$, and a puncturing pattern $\cP$ of size $m$.
 Instead of being transmitted on identical channels as in Ar{\i}kan's construction, the punctured bits, $x_i: i \in \cP$, are transmitted on zero capacity channels and the remaining
 $x_i$s are transmitted through $N'-m$ independent copies of $W$.  Let $y_i = 0$ for $i \in \cP$.
The modified channel $\tilde{W}_{N'}$, between the encoded information sequence $u_1^{N'}$ and the channel output $y_1^{N'}$, are described by the transitional probabilities
$\tilde{W}_{N'}(y_1^{N'}|u_1^{N'}) \triangleq  \prod_{i \in [N']\setminus \cP} W(y_i|x_i)$,
where $[N']$ is the set of indices $1,2,\dots,N'$.
Then, the modified bit-channels for $i \in [N']$ are given by
\begin{equation}
\begin{split}
\tilde{W}_{N'}^{(i)}(y_1^{N'}, u_1^{i-1}|u_i)
=\hspace{-5mm}\sum_{u_{i+1}^{N'}\in{\{0,1\}}^{N'-i}}\frac{1}{2^{N'-1}}\tilde{W}_{N'}(y_1^{N'}|u_1^{N'}).
\end{split}
\end{equation}
The sum-capacity can be found by generalizing \cite[lemma 4]{compound_polar}, where $\cC(W)$ denotes the symmetric capacity of channel $W$.

\begin{lemma}
\label{sum-capacity}
For any $m < N'$, $N' = 2^p$, and any puncturing pattern $\cP$ on the output of $G^{\otimes p}$ with cardinality $m$, the sum capacity is
$
\sum^{N'}_{i=1} \cC(\tilde{W}^{(i)}_{N'}) = (N' - m) \cC(W).
$
\end{lemma}
Hence, the portion of almost noiseless channels approach the average of the capacities of $\tilde{W}^{(i)}_{N'}$ which is equal to $(1 - \frac{m}{N'}) \cC(W)$ by \Lref{sum-capacity}. Applying the channel polarization theorem \cite{Arikan_09} and \cite[Theorem 5]{compound_polar}, we show that the two-step polarization of Fig. \ref{5kron7_g} results in a family of capacity-achieving punctured polar codes.

\begin{theorem}
Consider a polarization base of length $N'=2^p$ with $m$ punctured bits and $\beta < 0.5$.  There exists a family of two-step polarized codes approaching the rate $\mathcal{C}_m=(1-\frac{m}{N'})\cC(W)$ with vanishing probability of error bounded by $2^{-N^{\beta}}$, as code length $N$ goes to infinity.
\end{theorem}
Note that $ \cC_m$ is the average channel capacity after puncturing a fraction of $m/N'$ output bits.
The above theorem can be generalized to $\mathcal{C}_m=(1-\frac{m}{N'})\textbf{E}[\cC(W)]$  for the case of BICM channels, where $\textbf{E}[\cC(W)]$ is the average symmetric capacity of the BICM channels as described in subsection \ref{Sec_interleaver}.

\begin{figure}[t!]
\begin{center}
\includegraphics[width=0.48 \textwidth ]{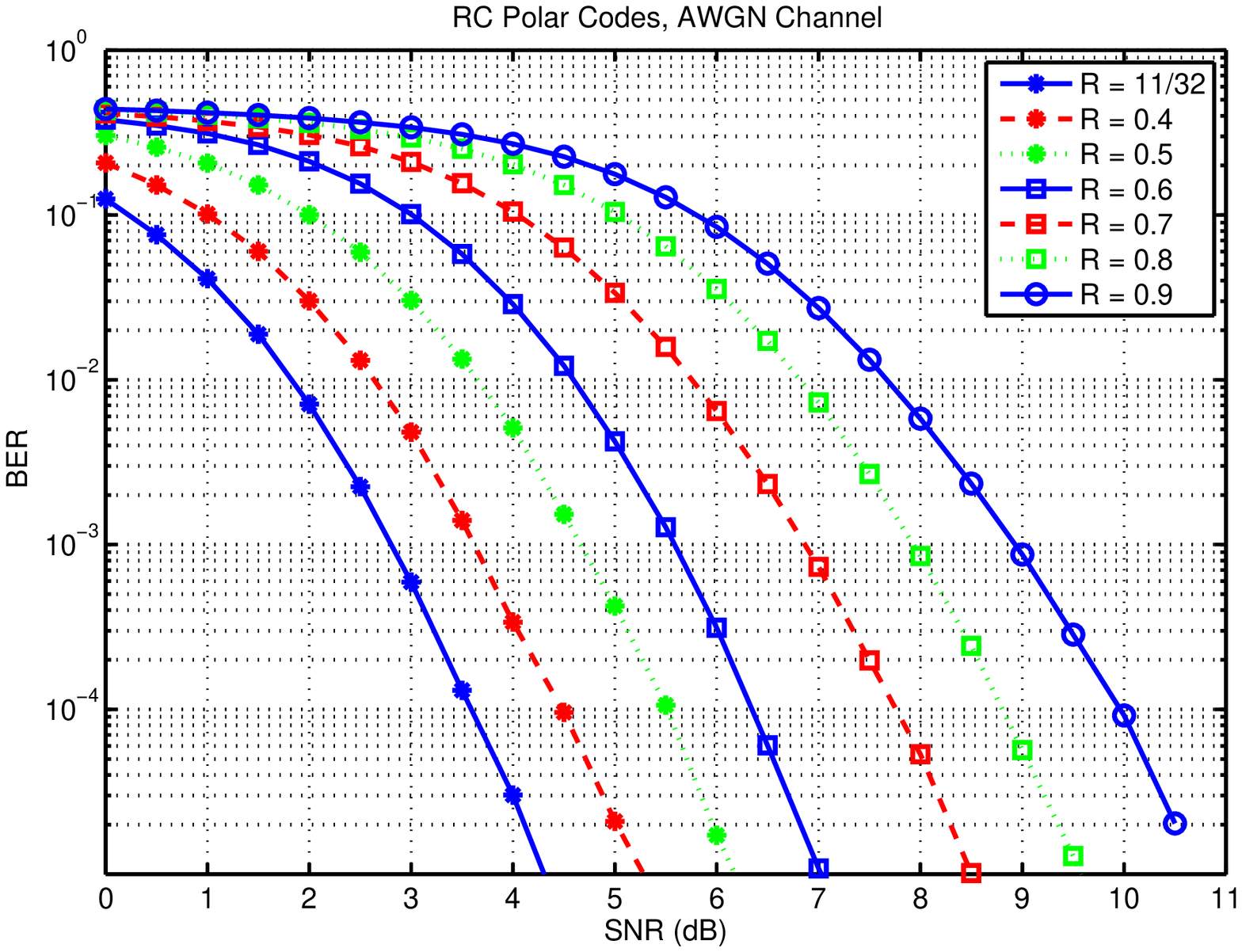}\\

\includegraphics[width=0.48 \textwidth ]{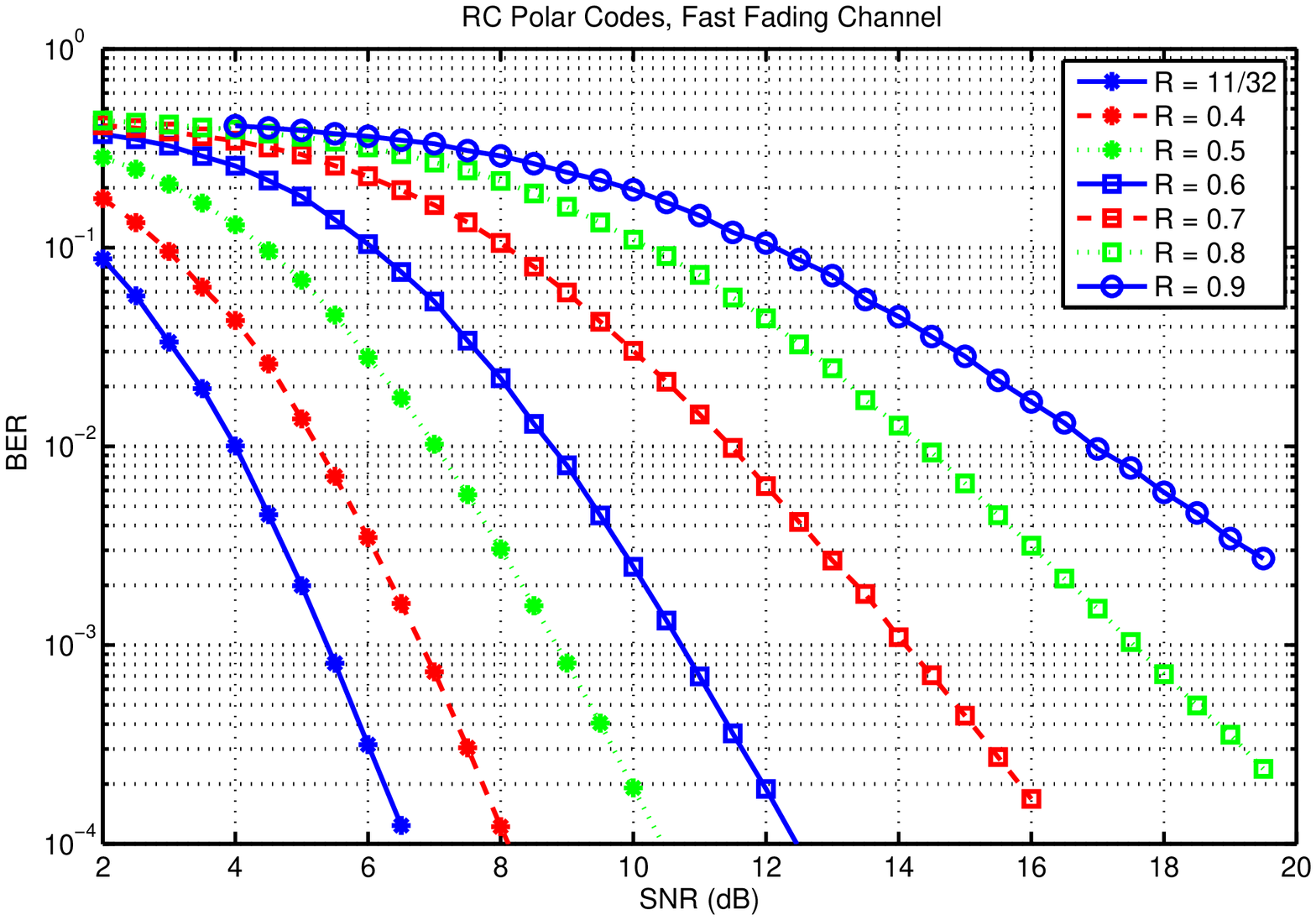}
\end{center}
\caption{BER of RC polar codes with  a fixed information set of size $88$ at different code rates under AWGN (top) and fast fading (bottom) channels \label{RC_256_fading}}
\vspace{-0.2cm}
\end{figure}

Next, we provide numerical simulations for the proposed interleaved RC polar codes with BICM. The progressive puncturing sequence is found on a base code of length $32$, from which two RC families with mother code lengths $N=256$ and $N=1024$ are constructed. Let $s_k$ be the $k$th transmitted QAM symbol, then the $k$th received symbol is $r_k=a_k s_k + n_k$, where $a_k$ is the zero-mean complex Gaussian fading coefficient or the Rayleigh fading coefficient for QAM or BPSK, respectively. $n_k$ is the additive white Gaussian noise (AWGN) with signal to noise ratio (SNR)
 $10 \log_{10}(1/(2 \sigma^2)$ in case of QAM, or $10 \log_{10}(1/\sigma^2)$ in case of BPSK. The puncturing sequence  $(0,\ 16,\ 8,\ 24,\ 2,\ 20,\ 26,\ 12,\ 10,\ 18,\ 4,\ 22,\ 25,\ 6,\ 13,\ 14,\ \\
1,\ 17,\ 28,\ 3,\ 5,\ 9,\ 29,\ 11,\ 19,\ 7,\ 21,\ 15,\ 23,\ 27,\ 30,\ 31)$ is found by running the PPA  with the GA at $\mbox{SNR}=3.5$ dB on the base code with rate $11/32$.

Fig. \ref{RC_256_fading} shows the bit error rate (BER) performance by SC decoding of the RC family with $N=256$ and $|\mathcal{I}|=88$ bits at code rates  $R \in \{0.9, 0.8, 0.7, 0.6, 0.5, 0.4, 11/32\}$   for BPSK transmissions on both AWGN and fast fading channels. To select the information set, the bit-channel error probabilities are estimated by genie-aided SC decoding of the codes with rate $R=88/98$ at SNRs of $3.5$ dB and $6.5$ dB  on AWGN and fading channels, respectively.

\begin{figure}[t!]
\begin{center}
\includegraphics[width=0.5 \textwidth ]{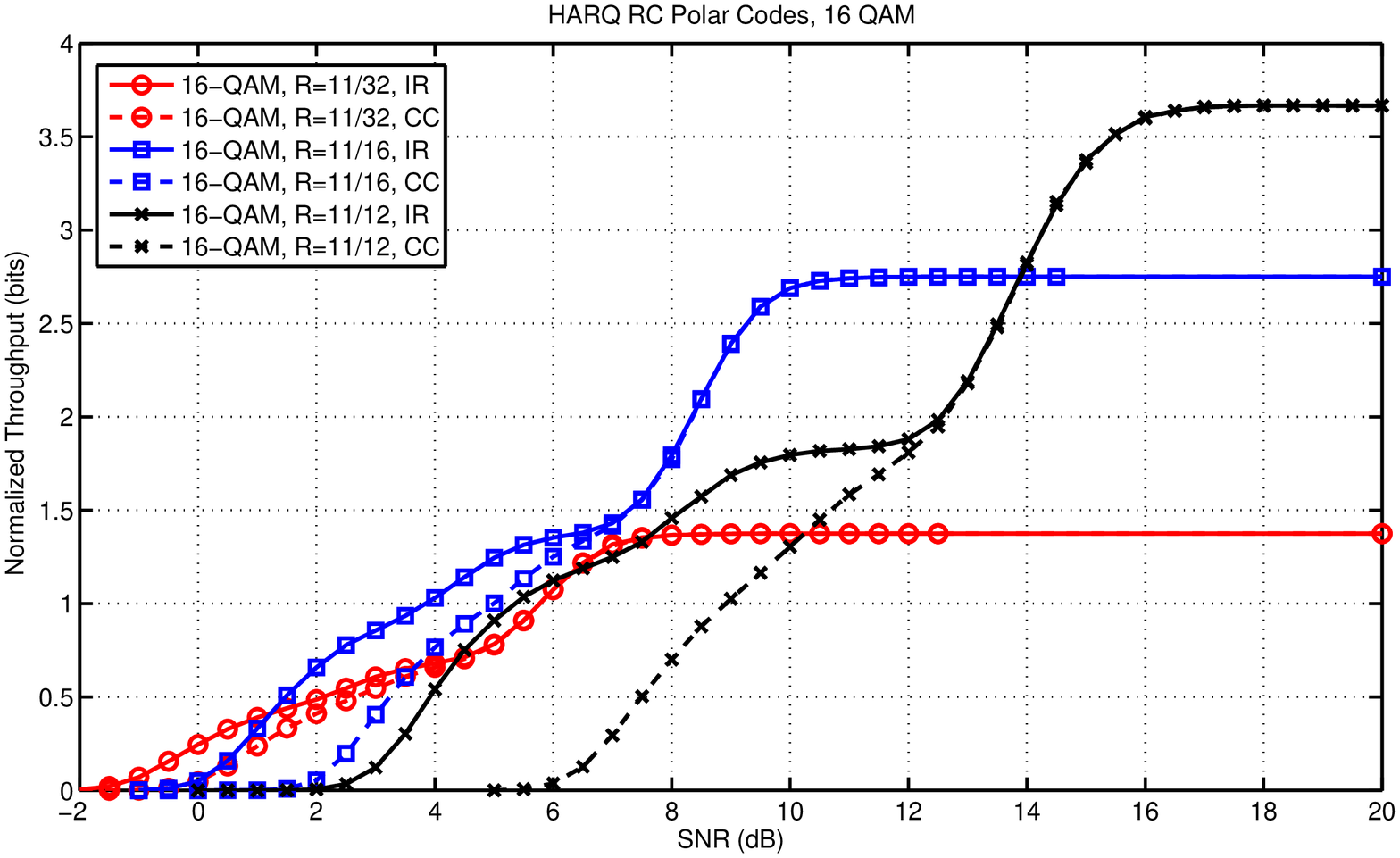} 
\\
\includegraphics[width=0.5 \textwidth ]{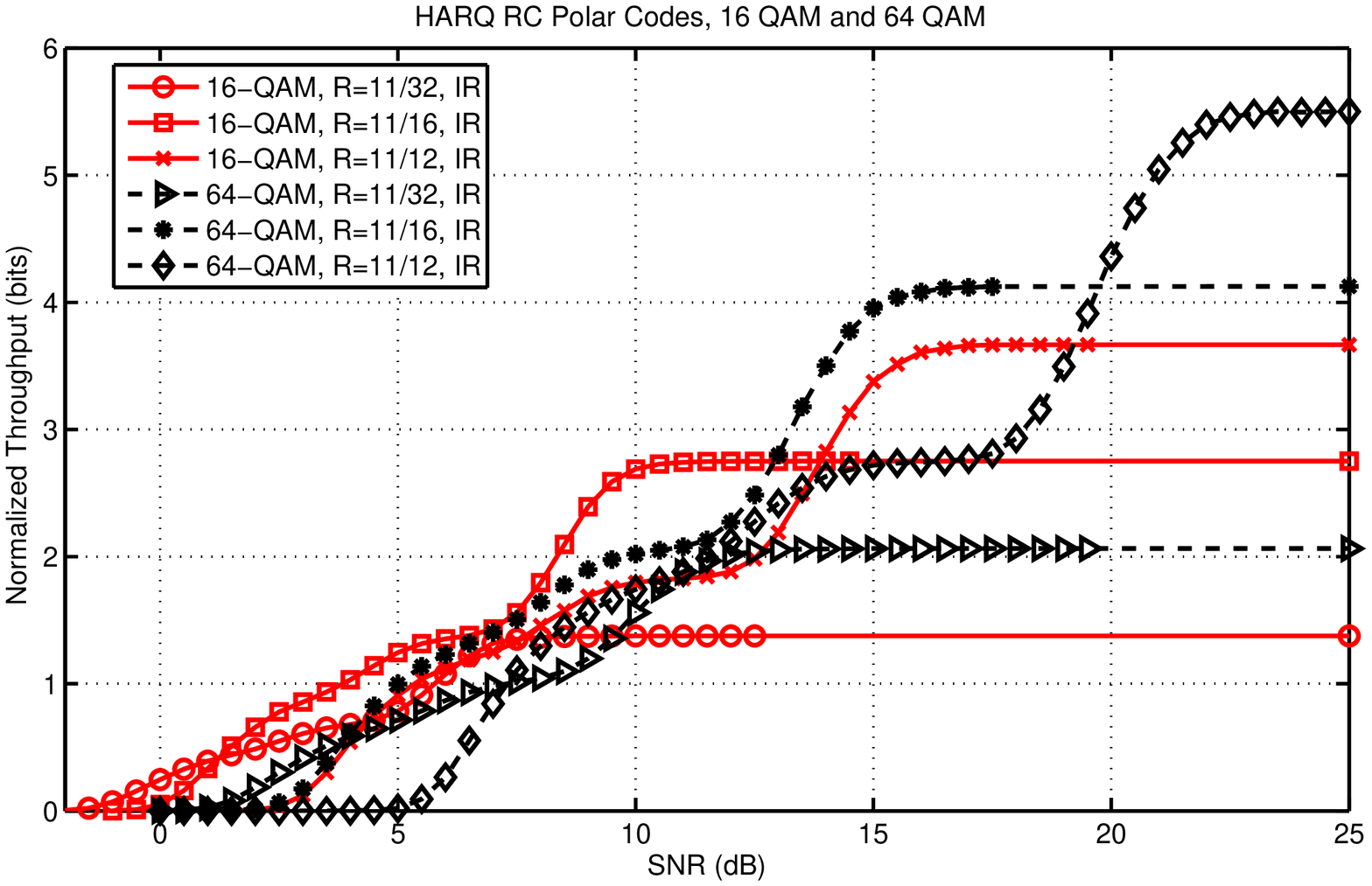}
\end{center}
\caption{Simulations of proposed HARQ RC polar codes: IR versus CC with 16-QAM (top)
and 16-QAM versus 64-QAM with IR (bottom).}
\vspace{-0.2cm}
\label{HARQ_1024_16qam}
\end{figure}

In Fig. \ref{HARQ_1024_16qam}, the normalized throughput with IR and CC HARQ schemes and BICM is illustrated
for the RC family with $N=1024$ and $|\mathcal{I}|=352$ bits.
A modulation and coding scheme (MCS) is defined by the code of rate $R$ and the modulation order $M$.
The \emph{normalized throughput} is defined as $\mathcal{T}=R\log_2(M)(1-\mbox{BLER})/\bar{t}$, where BLER is the average residual block error rate after all decoding attempts and $\bar{t}$ is the average number of transmissions required for successful decoding of an information block at the tested SNR.
Assuming a maximum of $t=4$ transmissions on an $16$-QAM AWGN channel, with 3 RC polar codes with rates from 0.34 to 0.92.  It is observed that IR provides about 3 dB gain over CC. By comparing IR and CC at $R=11/32$, we conclude that changing the bit-to-symbol mappings across transmissions provides extra diversity and better performance. In system design, one selects the MCS with the highest throughput at a given SNR. To maximize throughput,  it is best to utilize $16$-QAM with $R=11/32$ till an SNR of 1.2 dB, then use a transmission rate of $R=11/16$ till an SNR of $12.8$ dB, at which it is best to switch to $64$-QAM with $R=11/16$ up to SNR of 19.7 dB, then use the highest transmission rate $R=11/12$ with $64$-QAM  at higher SNRs.

\section{Conclusion}\label{conclude}

To construct families of rate-compatible polar codes, a  low-complexity algorithm that progressively selects the puncturing order on a base code of short length is devised and shown to have near-optimal performance.
Based on a two-step polarization construction, a flexible rate-matching scheme selects the transmitted bits for HARQ transmissions.
Whereas previous algorithms would require storing the puncturing patterns for each designed code rate and length, the proposed scheme only needs to store one short (\emph{e.g.} of length 32) puncturing sequence, which will be used to generate the puncturing pattern at any desired code rate and length.
The proposed rate-matching and bit-mapping scheme  preserves the code polarization in the presence of punctured bits and multi-channel bit-interleaved coded bits, and can result in any code rate by either puncturing or repetition of the coded bits in a particular order.  IR and CC HARQ schemes with adaptive bit-interleaved polar-coded modulation on wireless channels were investigated. It is  shown in this paper that punctured polar codes can have a capacity achieving property.

Furthermore, the proposed scheme can be applied with other polar code constructions such as the reduced-complexity relaxed polar code constructions \cite{relaxed_polar_code} or  concatenated constructions that have improved burst-error correction capability \cite{mahdavifar2014performance}. With concatenated decoding \cite{mahdavifar2014performance} or list decoding \cite{FastPolar}, the performance of RC polar codes is expected to be comparable to those of RC turbo codes and RC LDPC codes, \emph{cf.} \cite{el2009design}, without suffering from error floors. The polar code rate-matching and bit-mapping schemes of this paper make the adoption of polar codes in future wireless systems more practical.

\bibliographystyle{IEEEtran}
\bibliography{RCbib2}

\end{document}